# Electrical probing of COVID-19 spike protein receptor binding domain via a graphene field-effect transistor

Xiaoyan Zhang[a,d], Qige Qi[a], Qiushi Jing[a], Shen Ao[a,b], Zhihong Zhang[c], Mingchao Ding[c], Muhong Wu[c], Kaihui Liu[c], Weipeng Wang[b], Yunhan Ling[b], Zhengjun Zhang[b], Wangyang Fu[a,b,*]

[a] State Key Laboratory of New Ceramics and Fine Processing, School of Materials Science and Engineering, Tsinghua University, Shaw technical science building, Haidian District, Beijing, 100084, P. R. China

[b] Key Laboratory of Advanced Materials of Ministry of Education, School of Materials Science and Engineering, Tsinghua University, Shaw technical science building, Haidian District, Beijing, 100084, P. R. China

[c] State Key Laboratory for Mesoscopic Physics, Frontiers Science Center for Nano-optoelectronics, School of Physics, Peking University, Beijing 100871, China

[d] Leiden Institute of Chemistry, Leiden University, Einsteinweg 55, 2333CC Leiden, the Netherlands

*Correspondence author: fwy2018@mail.tsinghua.edu.cn

## Abstract

Here, in an effort towards facile and fast screening/diagnosis of novel coronavirus disease 2019 (COVID-19), we combined the unprecedently sensitive graphene field-effect transistor (Gr-FET) with highly selective antibody-antigen interaction to develop a coronavirus immunosensor. The Gr-FET immunosensors can rapidly identify (about 2 mins) and accurately capture the COVID-19 spike protein S1 (which contains a receptor binding domain, RBD) at a limit of detection down to 0.2 pM, in a real-time and label-free manner. Further results ensure that the Gr-FET immunosensors can be promisingly applied to screen for high-affinity antibodies (with binding constant up to $2\times10^{11}$ $M^{-1}$ against the RBD) at concentrations down to 0.1 pM. Thus, our developed electrical Gr-FET immunosensors provide an appealing alternative to address the early screening/diagnosis as well as the analysis and rational design of neutralizing-antibody locking methods of this ongoing public health crisis.

Rapid and accurate identification/characterization of a potential pathogen is crucial for disease control, patient treatment and prevention of epidemic of emerging infectious diseases, such as the severe acute respiratory syndrome coronavirus (SARS-COV), which has incurred pandemics of respiratory infectious diseases with high mortality. Recently, another emerging coronavirus that can cause viral pneumonia (COVID-19) is outbreak in Wuhan, China and has propagated across the entire China. Meanwhile it has already shown great potential to have globe spread and thread to the worldwide public health.[1, 2] To date, nucleic acid-based molecular diagnostic tests based on the reverse transcription-polymerase chain reaction (RT-PCR),[3] have been established and widely adopted to identify this COVID-19. However, these detection methods require sophisticated primer and probe design, multi-step reactions, many reagents, trained personnel and bulky instruments. Moreover, it is impossible or inaccurate to detect infected but recovered people or asymptomatic carriers. In this respect, simple and cost-efficient protein-based immunosensor with high sensitivity is of vital importance when pestilential blast.

Graphene field-effect transistors (Gr-FET) are very attractive for the development of immunosensors due to their unprecedented sensitivity and chemical stability with proved capability for label-free digital biomolecules detection.[4-6] Compared to optical based immunoassays including fluorescence-linked and/or enzyme-linked immunosorbent assay (FLISA and/or ELISA) technology, Gr-FET biosensors eliminate the complicated procedure for fluorescence or enzyme labeling and do not require bulky and expensive optical instruments. Thus, facile Gr-FET biosensors with high sensitivity hold great potential for rapid diagnosis and early risk prediction of infectious diseases, which often present low copy numbers in vivo pathogen. In the case of coronaviruses, it has been reported that its spike (S) glycoprotein plays an extremely important role in recognizing the cell surface receptor, which is essential for both host specificity and viral infectivity. Further studies confirms that COVID-19 infects the human respiratory epithelial cells through the S1 subunit protein, which mainly contains a receptor binding domain (RBD) interacting with the human angiotensin-converting enzyme 2 (ACE2).[7] Moreover, the S protein also plays key parts in the induction of neutralizing-antibody and T-cell responses, as well as protective immunity.[8]

Here, to enable digital detection of the virion attachment protein of coronaviruses for fast screening/diagnosis, we combined the extremely sensitive Gr-FET with highly selective SARS-COV spike S1 subunit protein antibody (CSAb) - COVID-19 spike S1 subunit protein (which contains the RBD) antigen interaction to develop an immunosensor. Strikingly, CSAb modified Gr-FET is capable of real-time detecting S1 at a limit of detection (LOM) down to 0.2 pM concentration with a fast responding time within 2 mins. Moreover, our results indicate that CSAb modified Gr-FET possesses a higher sensitivity than its counterpart with ACE2 receptors, which can be attributed to the higher bonding affinity of CSAb to S1 ($K$=2×$10^{11}$ $M^{-1}$) compared to that of ACE2 ($K$=$10^9$ $M^{-1}$). Thus, along with the capability of utilizing S1 modified Gr-FET to screening antibodies, we expand the potential utility of Gr-FET for rapid screening/diagnosis of respiratory infections caused by coronaviruses and its fast drug screening.

We started with high-quality monolayer single crystal graphene synthesized on single crystal Cu(111) foils by a chemical vapor deposition (CVD) method (Fig. S1),[9] and fabricated high-performance, solution-compatible Gr-FETs following an "upside-down" process described previously.[10] Figure 1a depicts the schematics of Gr-FET for coronaviruses diagnosis, where graphene surface was specifically functionalized with either CSAb or ACE2 receptor to bind the S1 subunit protein from COVID-19. The hybridization of the slightly positively charged S1 protein with CSAb/ACE2 receptors immobilized on the graphene surface, alters its conductance/resistance via field effect, which can be electrically read out. We note here that we applied a reference electrode in constant contact with the antigen buffer solution to fix its electrostatic potential ($V_{ref}$) during antibody-antigen reaction, and to control the current flow in the graphene channel between the source and the drain electrodes. As CSAb is positively charged, while ACE2 is negatively charged in PBS buffer solution (pH=7.2), we applied a negative or a positive potential at graphene during the incubation process, respectively, to improve their immobilization on the sensor surface.

Figure 1b plots the sheet conductance ($G$) of graphene measured at a constant current source $I_{ds}$ ~ 1 μA when sweeping the liquid-gate voltage $V_{ref}$ ranging from −0.5 to 0.3 V. The transfer characteristics gives the liquid-gate voltage of a bare Gr-FET device at the charge neutrality point (CNP, also called Dirac point) $V_{Dirac}$=-90 mV, where the total charge carrier number is minimal. Initial field-effect carrier mobilities can be deduced as 1017 $cm^2$ $V^{-1}$ for hole and 1215 $cm^2$ $V^{-1}$ for electron. 30 min-incubation of CSAb results in a shift of the Dirac point $\Delta V_{Dirac}$=-50 mV towards negative direction (red dots, Fig. 1b), which can be attributed to the positive charges carried by the immobilized antibodies. To confirm the adhesion of the self-assembly CSAb on graphene surface, in another Gr-FET sample we treated the CSAb immobilized sensor surface with tween 20 (T20), a surfactant with known blocking/passivation effect on graphene surface. We observed negligible change in the transfer characteristics before and after T20 passivation for 30 mins (Fig. S2), suggesting its good adhesion and tight surface coverage via hydrophobic and electrostatic interactions during its self-assembly on graphene surface.

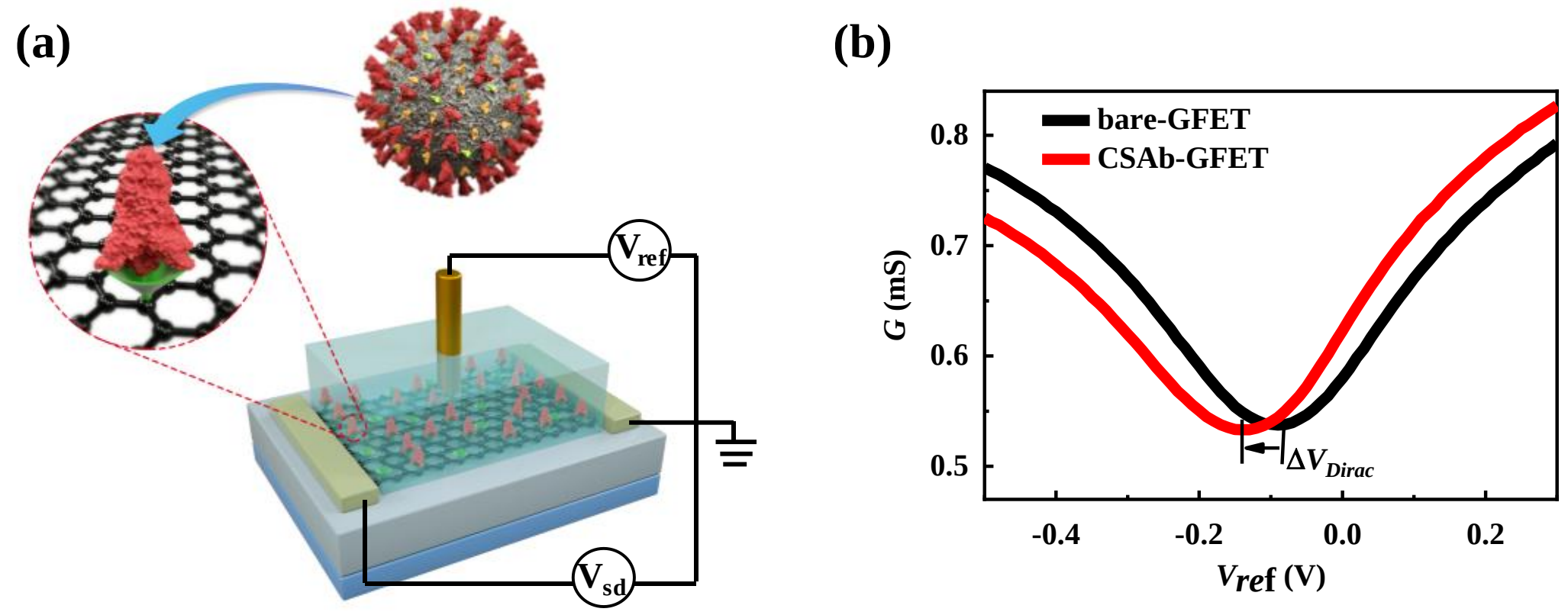

**Figure 1.** Protein functionalized Gr-FET immunosensors. (a) Schematics of the Gr-FET immunosensors. We functionalize the graphene surface with either CSAb or ACE2 receptor, which have proved affinity for the S1 subunit protein (which contains the RBD) from coronaviruses. (b) Electrical source-drain sheet conductance $G$ as a function of the reference potential $V_{ref}$ measured for the CSAb immobilized Gr-FET. A bipolar transfer curve is observed corresponding to different type of charge carriers that can continuously be tuned from holes (left) to electrons (right) with a Dirac point $V_{Dirac}$ at minimum $G$.

Due to Debye screen in the 10 mM PBS buffer solution (Debye length=0.7 nm), the charged S1 subunit protein cannot approach the graphene surface close enough to induce a notable field-effect, as the expected size of the antibody is around 7-10 nm,[11] i.e., one order-of-magnitude larger than the Debye length of the buffer solution. To alleviate the ionic screening effect, we performed all the real-time sensing test under diluted 0.01 PBS with elongated Debye length of 7 nm (comparable to the size of the CSAb).

During real-time antibody-antigen reaction, we fixed the liquid-gate voltage ($V_{ref}$) at 0 V (electron branch), where the $G$-$V_{ref}$ curve exhibit a sharp slope, thus a high sensing response. For signal processing, we normalized the change of the conductance $\Delta G$ by the transconductance $g_m$ to deduce the shift of the liquid-gate voltage $\Delta V_{ref}=\Delta G/g_m$ of Gr-FET upon antibody-antigen reaction. Such normalization helps to eliminate the effect of sensor-to-sensor variation in electrical parameters, thus yielding a reliable and reproducible sensing performance. Figure 2a shows the real-time responses ($\Delta V_{ref}$) monitored at varying concentrations of S1 ranging from 0.2 pM to 10 nM (added at time intervals of about 5 mins with increasing concentration). The decrease of $\Delta V_{ref}$ depending on the increasing concentration of S1 is consistent with the bonding of slightly positively charged molecules on the sensor surface. Such CSAb-S1 complex can be refreshed and reused with comparable performance after thoroughly rinsing with buffer solution (Fig. S3). Additionally, repeating the same measurement when shifting the liquid-gate voltage from $V_{ref}$=0 V (electron branch) to $V_{ref}$=-0.2 V (hole branch), yields similar sensing behavior as expected (Fig. S4). As a negative control, pure T20 modified Gr-FET was also tested (Fig. S5). Compared to the CSAb immobilized Gr-FET immunosensor, T20 passivated Gr-FET shows negligible sensing response to the target S1. Therefore, our results clearly suggest that Gr-FET signal output is specific to the immobilized CSAb-S1 complex. Strikingly, such CSAb immobilized Gr-FET immunosensors target S1 subunit protein of COVID-19 with a limit of detection (LOD) down to 0.2 pM, which rivals that of state-of-the-art ELISA technology but eliminating the requirements of complicated procedure for enzyme labeling or bulky/expensive optical instruments.[12, 13] We note here that there is still plenty of room for improving the LOD by innovative antibody design and/or optimal Gr-FET sensing scheme. In addition, the real-time monitoring of the $\Delta V_{ref}$ response (Fig. 2a) indicates a fast response time (within 2 min) for the detection of S1, which competes the current fast FLISA and/or ELISA technologies.[13]

Alternatively, ACE2 is an integral membrane protein served as functional receptor for the spike glycoprotein of human coronaviruses SARS and COVID-19 infections. We further examined the affinity of ACE2-functionalized Gr-FET in the presence of S1 subunit protein at varying concentrations (10 pM–1 μM). In contrast to CSAb, real-time measurement of ACE2-functionalized Gr-FET immunosensor (Fig. 2b) suggest that sharp increase/change of $\Delta V_{Dirac}$ happened only when 1 nM S1 was added. Further affinity fitting based on a single-reaction model[10] (see also SI) in Fig. 1e give binding affinities of $K$=2×10$^{11}$ M$^{-1}$ and $K$=10$^{9}$ M$^{-1}$ for CSAb-S1 and ACE2-S1 interaction, respectively. Obviously, compared to the ACE2 modified Gr-FET, its CSAb modified counterpart exhibits a much higher affinity to the S1 subunit protein of COVID-19,[14, 15] which is in consistent with previous reports that the antibody can scavenging virus before they bonding to receptors.

Neutralizing antibodies are extraordinarily crucial in the development of vaccines and antibody drugs. Here, to screen the neutralizing antibody, we functionalized Gr-FET with S1 protein to differential its affinity to CSAb or ACE2 in real-time. The results indicate a sensing response to CSAb even at 0.1 pM concentration;(Fig. 2d) whereas clear and sharp response can be identified for ACE2 only after 1 nM (Fig. 2e). In Fig. 1f, affinity fitting using our previously determined binding constants of $K$=2×10$^{11}$ M$^{-1}$ and $K$=10$^{9}$ M$^{-1}$ gives good description for CSAb-S1 and ACE2-S1 interactions, respectively, also confirming our previous results and conclusions on CSAb or ACE2 immobilized Gr-FET for S1 detection. Here we ascribe the slight deviations in the affinity fitting for CSAb-S1 interaction at 0.1 pM and 1 pM concentration (Fig. 2f) to possible ionic screening of the weakly charged CSAb with relatively large size.

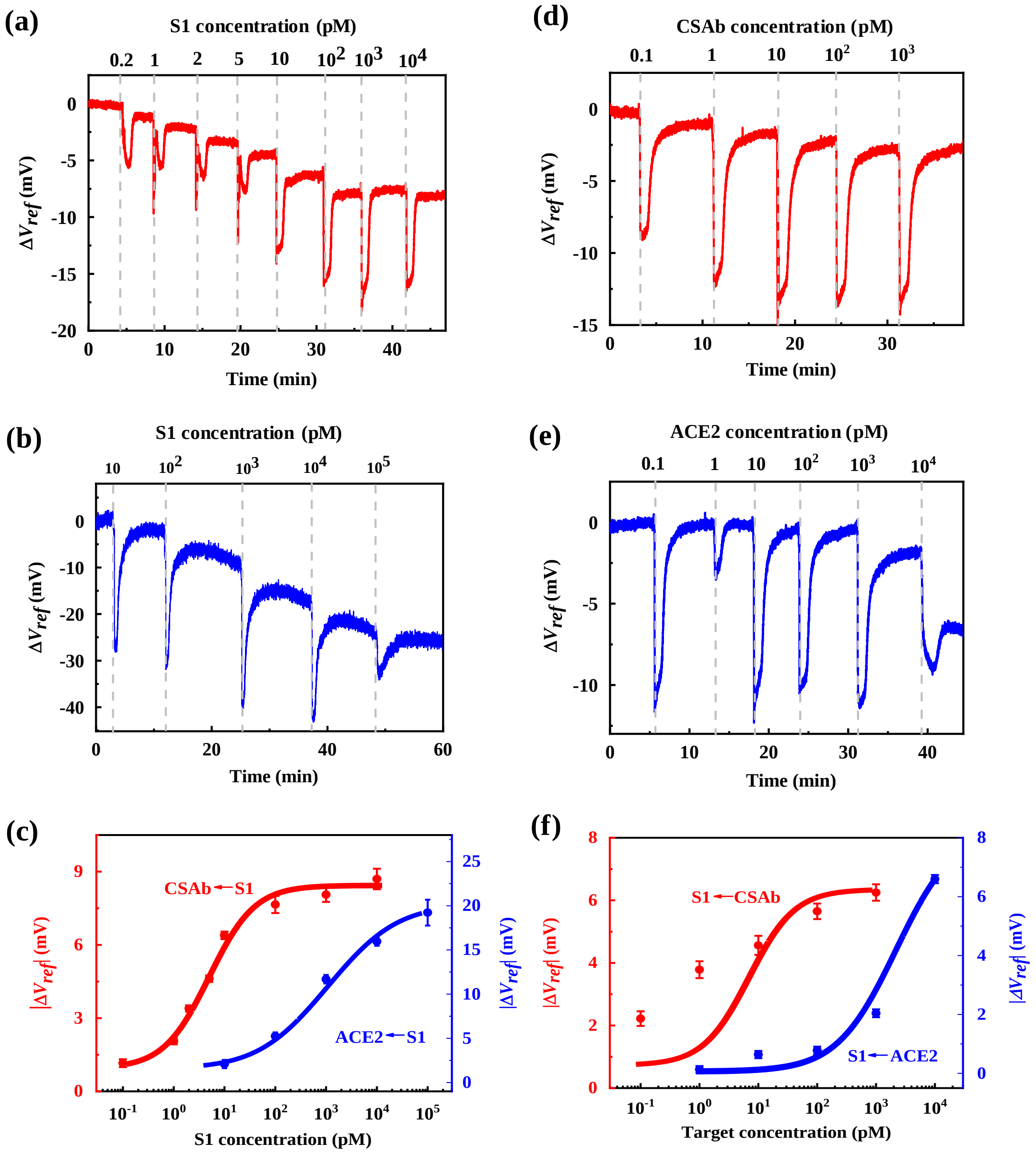

**Figure 2.** Protein functionalized Gr-FET immunosensors for detection of COVID-19 related proteins. (a) Changes of the CSAb immobilized Gr-FET's $V_{ref}$ versus time upon the introduction of S1 solutions at various concentration from 0.2 pM to 10 nM in 0.1 mM PBS buffer solution. (b) Changes in $V_{ref}$ of ACE2 immobilized Gr-FET versus time upon the introduction of S1 solutions with various concentration from 10 pM to 1 μM in 0.1 mM PBS buffer solution. (c) Affinity fittings of the CSAb and the ACE2 mobilized Gr-FETs for CSAb-S1 (red dot and red line fit) and ACE2-S1 (blue dot and blue line fit) interactions, yield binding affinities $K=2\times10^{11}$ $M^{-1}$ and $K=10^{9}$ $M^{-1}$, respectively. (d) Detection of CSAb with S1 spike functionalized Gr-FET biosensors. Changes of the S1 immobilized Gr-FET's $V_{ref}$ versus time upon the introduction of CSAb solutions at various concentrations ranging from 0.1 pM to 10 nM in 0.1 mM PBS buffer solution. (e) Detection of ACE2 with S1 spike functionalized Gr-FET biosensors. Changes of the CSAb immobilized Gr-FET's $V_{ref}$ versus time upon the introduction of ACE2 solutions at various concentrations ranging from 0.1 pM to 10 nM in 0.1 mM PBS buffer solution. (f) Affinity fittings using previously obtained equilibrium constants of S1 immobilized Gr-FET for CSAb-S1 (red dot and red line fit) and ACE2-S1 (blue dot and blue line fit) interactions, respectively.

In summary, we expand the applications of label-free Gr-FET digital biosensing technology to fast detection of COVID-19 spike protein S1 (which contains the RBD) with sensitivities rivaling those of state-of-the-art ELISA technologies, but eliminating any requirements of complicated procedure for enzyme labeling or bulky/expensive optical instruments. We constructed the Gr-FET immunosensors by immobilizing either CSAb or ACE2 on the surface of the graphene, both of which bind specifically to the spike protein RBD. A direct comparison between CSAb and ACE2 based Gr-FET biosensors indicates that CSAb has a higher affinity to the spike protein RBD with a LOM down to 0.2 pM. Moreover, we demonstrated the principle of spike protein S1 antigen functionalized Gr-FETs for fast analysis and screening of neutralizing antibodies, which can block coronaviruses from attaching and infecting the health cell. Thus, these results have important implications in the context of rapid and facile digital immune-specific identification, as well as research and development of vaccines, prophylactics and therapeutics to combat COVID-19 and other emerging coronavirus associated infectious diseases, specifically during the outbreak period of pandemics.

## Acknowledgments

The authors acknowledge financial support from the National Natural Science Foundation of China (51991342), the Swiss National Science Foundation, European Commission Horizon 2020-Research and Innovation Framework Programme (Marie Sklodowska-Curie actions Individual Fellowship No.749671). Beijing Natural Science Foundation (JQ19004), Beijing Excellent Talents Training Support (2017000026833ZK11), The Key R&D Program of Guangdong Province (2019B010931001), Bureau of Industry and Information Technology of Shenzhen (No. 201901161512)

## Conflict of interest

All authors declare that they have no conflict of interest.

## Author contributions

Xiaoyan Zhang and Wangyang Fu designed the experiments. Xiaoyan Zhang prepared the Gr-FET devices. Wangyang Fu performed the electrical measurements. Zhihong Zhang, Mingchao Ding, Muhong Wu and Kaihui Liu provided the CVD graphene and carried related characterization. Qige Qi, Qiushi Jing and Shen Ao participated in the discussion and analysis on experimental results. Xiaoyan Zhang and Wangyang Fu wrote the manuscript. Kaihui Liu, Weipeng Wang, Yunhan Lin and Zhengjun Zhang contributed to the editing and revision of the manuscript. All authors read and approved the final manuscript.

**Electronic supplementary materials**

# Electrical probing of COVID-19 spike protein receptor binding domain via a graphene field-effect transistor

Xiaoyan Zhang[a,d], Qige Qi[a], Qiushi Jing[a], Shen Ao[a,b], Zhihong Zhang[c], Mingchao Ding[c], Muhong Wu[c], Kaihui Liu[c], Weipeng Wang[b], Yunhan Ling[b], Zhengjun Zhang[b], Wangyang Fu[a,b,*]

[a] State Key Laboratory of New Ceramics and Fine Processing, School of Materials Science and Engineering, Tsinghua University, Shaw technical science building, Haidian District, Beijing, 100084, P. R. China

[b] Key Laboratory of Advanced Materials of Ministry of Education, School of Materials Science and Engineering, Tsinghua University, Shaw technical science building, Haidian District, Beijing, 100084, P. R. China

[c] State Key Laboratory for Mesoscopic Physics, Frontiers Science Center for Nano-optoelectronics, School of Physics, Peking University, Beijing 100871, China

[d] Leiden Institute of Chemistry, Leiden University, Einsteinweg 55, 2333CC Leiden, the Netherlands

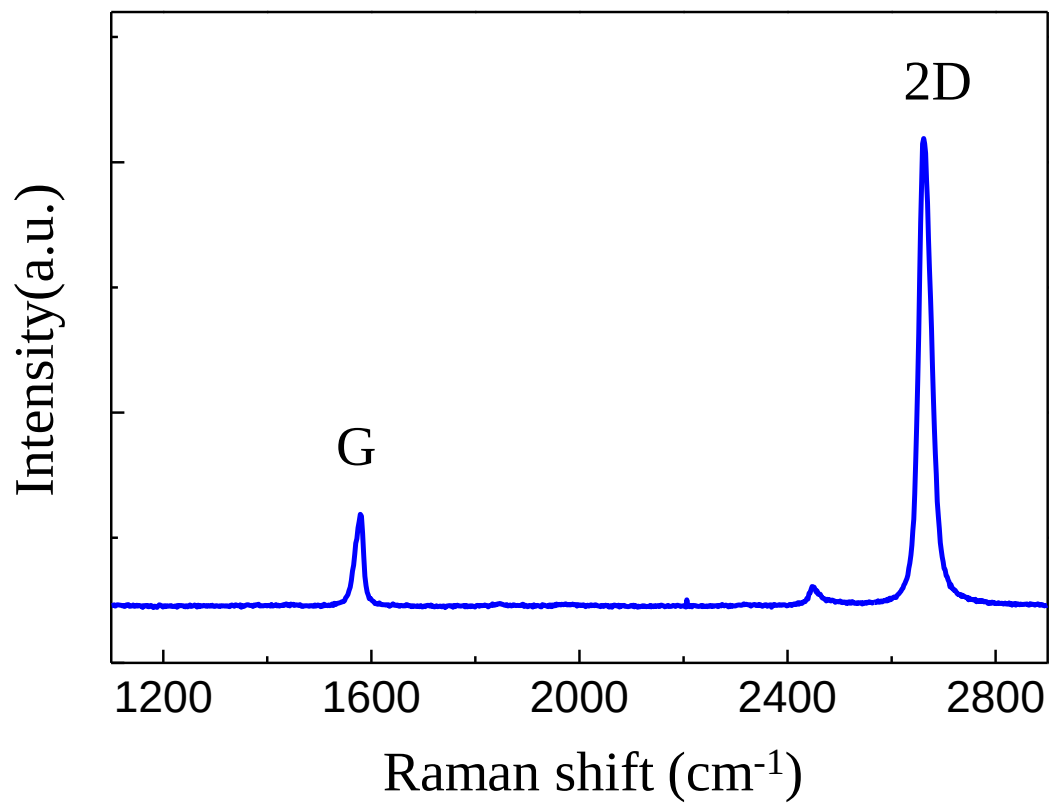

Fig. S1 Raman spectroscopy of monolayer single crystal CVD graphene for fabrication of Gr-FET immunosensors.

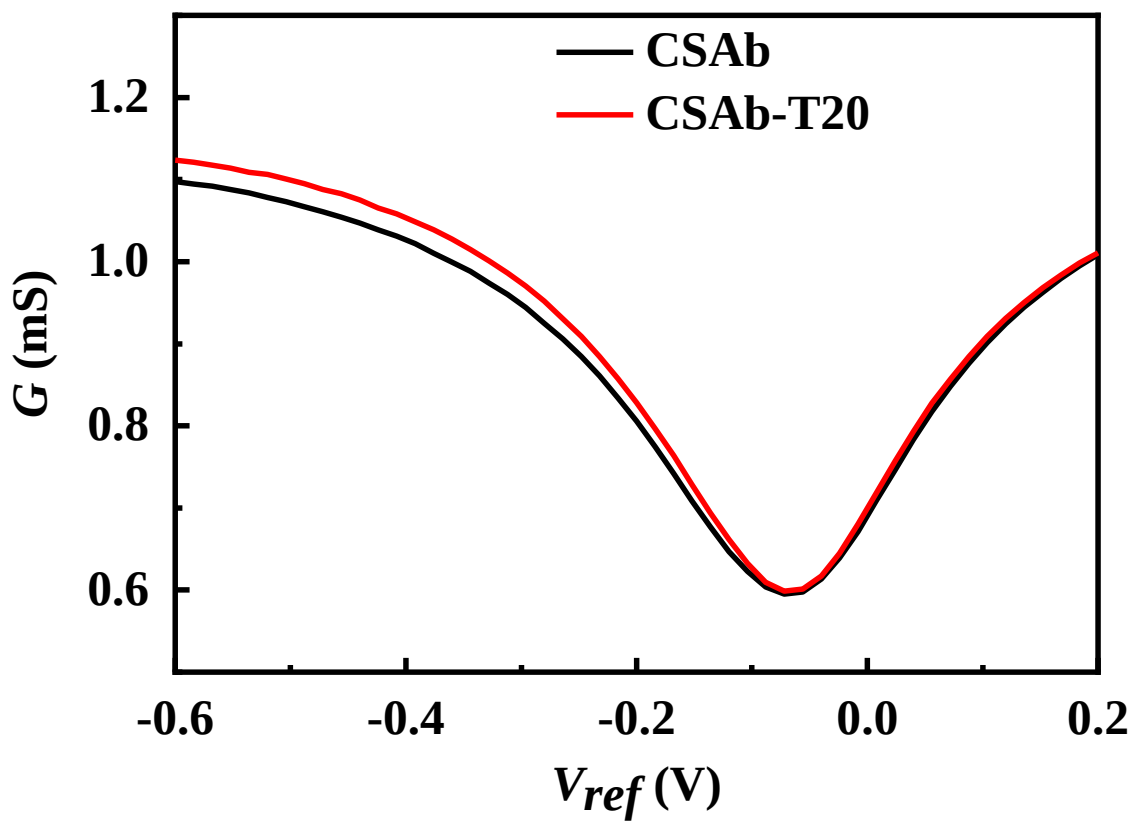

Fig. S2 Electrical source-drain conductance *G* as a function of the reference voltage $V_{ref}$ measured for the CSAb-immobilized Gr-FET before and after surface passivation by 30-min-T20 incubation at 0.05 wt% concentration. The results indicate a negligible change on the Gr-FET electrical performance due to T20 passivation.

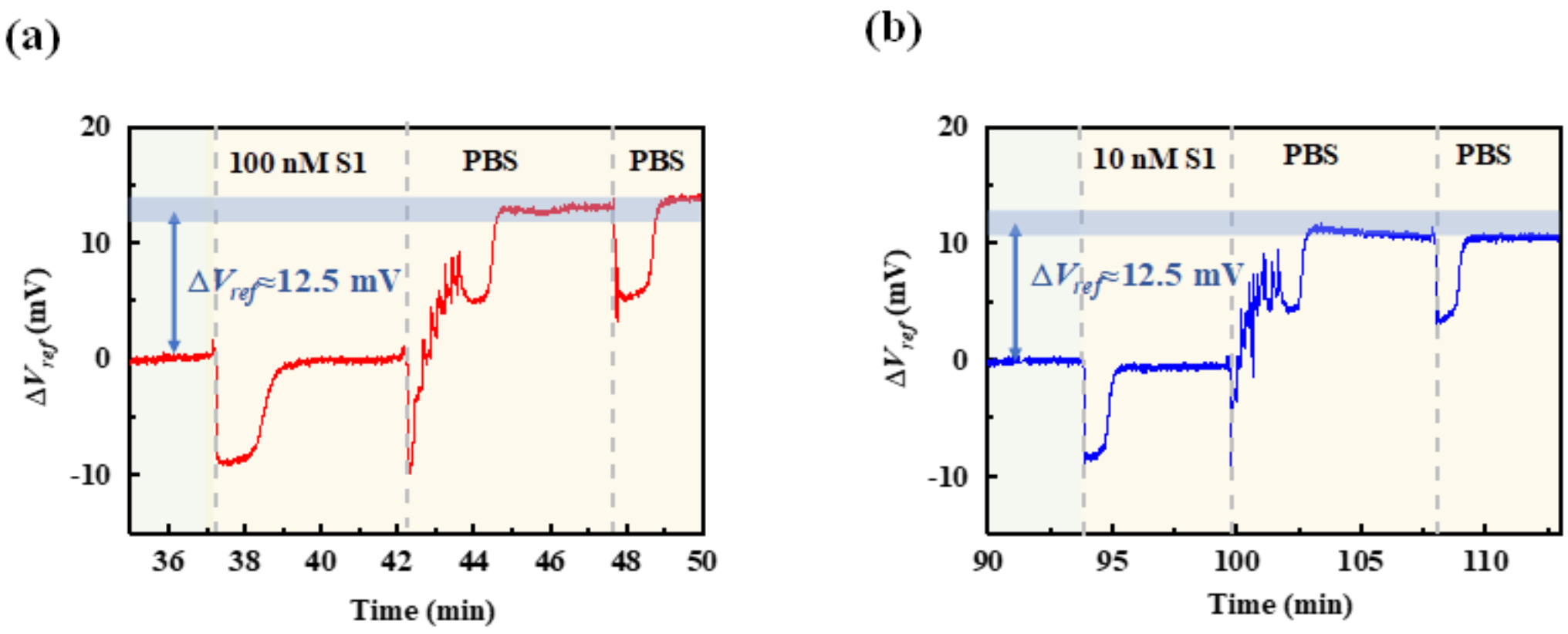

Fig. S3 Changes of the liquid-gate voltage $\Delta V_{ref}$ upon sufficient buffer solution rinsing suggests that CSAb-S1 complex can be refreshed. (a) First time refreshment and (b) second time refreshment. $\Delta V_{ref}$≈12.5 mV indicates the baseline level of the CSAb modified Gr-FET.

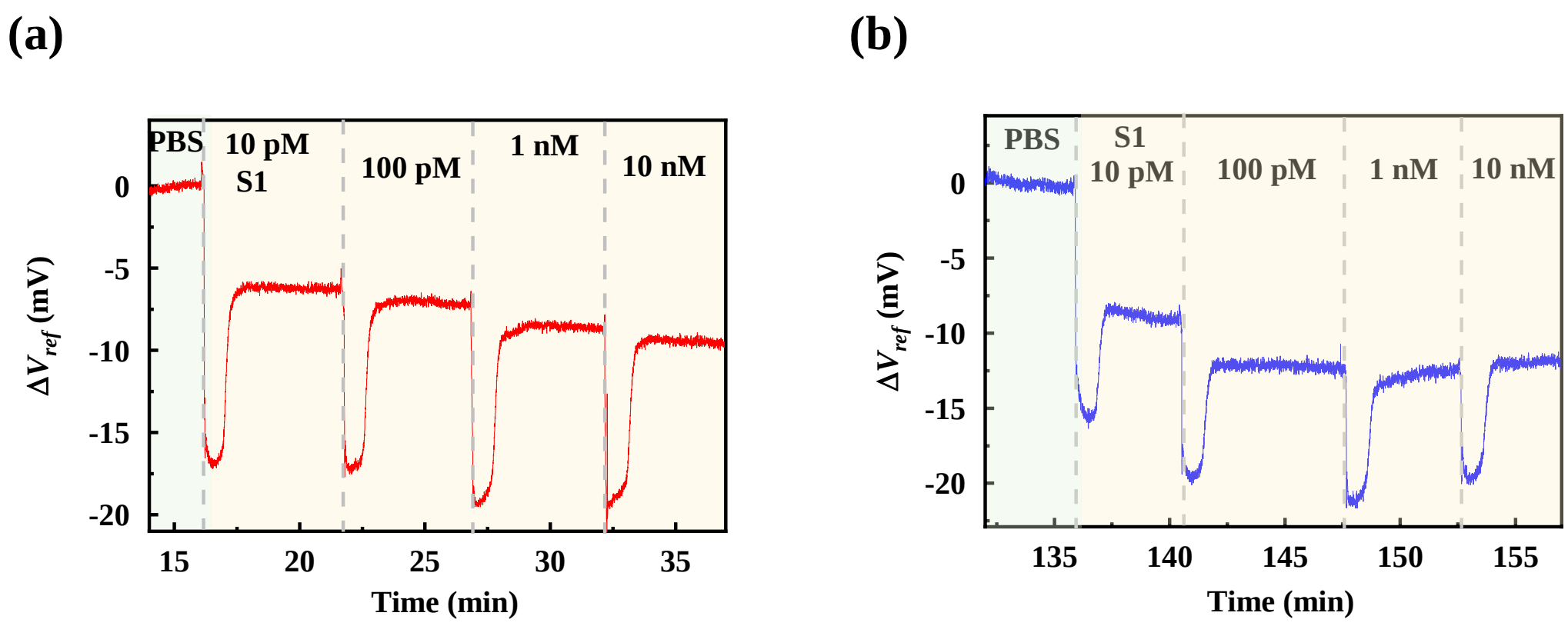

Fig. S4 (a) Change of the liquid-gate voltage $\Delta V_{ref}$ of CSAb immobilized Gr-FET versus time upon the introduction of S1 solutions at various concentration from 10 pM to 10 nM. (b) Shifting the liquid-gate voltage from $V_{ref}$=0 V (a) to $V_{ref}$=-0.2 V (and fixing at -0.2 V) yields similar sensing behavior for CSAb immobilized Gr-FET upon the introduction of S1 solutions at various concentrations ranging from 10 pM to 10 nM.

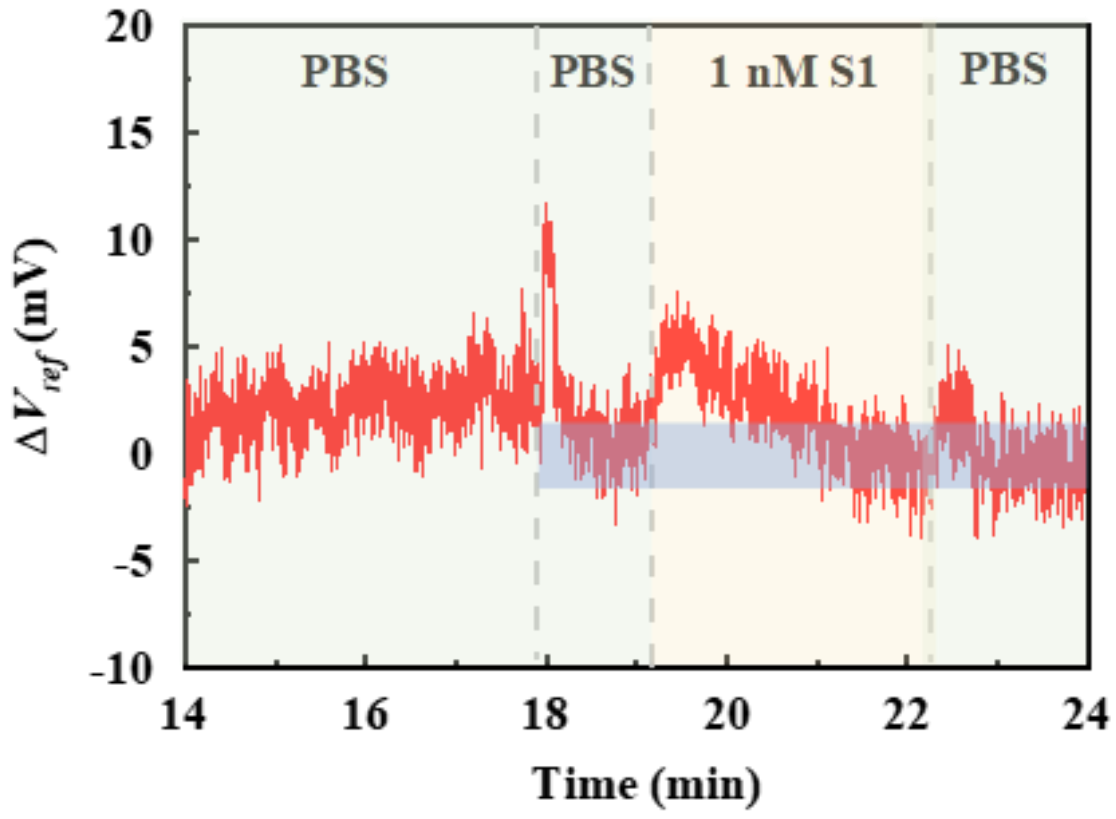

Fig. S5 Change of the gate voltage of T20 passivated Gr-FET versus time upon the introduction of S1 solution at 1 nM. No clear sensing response can be observed.

**Single-reaction model**

The $CSAb - S1$ complex system can be expressed by the dissociation constant $K_d$:

$$CSAb - S1^{x+} \overset{K_d}{\Leftrightarrow} CSAb + S1^{x+}, K_d = \frac{v_{CSAb} \cdot a_{s1^{x+}}}{v_{CSAb-S1^{x+}}} = \frac{1}{K} \quad \text{(S1)}$$

here $v$ is the number of sites of each species, $a_{S1^{x+}}$ the activity of the positively charged $S1^{x+}$ at the sensor surface, $K$ the binding affinity. The active surface sites are either complex $CSAb - S1^{x+}$, or free $CSAb$. The total number of these sites is:

$$N_s = v_{CSAb-S1^{x+}} + v_{CSAb} \quad \text{(S2)}$$

For simplifying the calculation, we assume the $CSAb$ and $S1$ carrier zero and $x$e charges, respectively. The complex $CSAb - S1^{x+}$ generates the total surface charge:

$$\sigma_0 = xev_{CSAb-S1^{x+}} \quad \text{(S3)}$$

Eq. S1-3 can be rewritten:

$$\sigma_0 = xeN_s(\frac{a_{S1^{x+}}}{a_{S1^{x+}} + K_d}) \quad \text{(S4)}$$

These surface charge $\sigma_0$ are screened by the ions of the double layer, yielding a surface potential drop $\Psi_0$ over the double-layer capacitance $C_{DL}$:

$$\sigma_0 = C_{DL} \cdot \Psi_0 \quad \text{(S5)}$$

Via the Boltzmann equation, we relate the (unknown) activity of the surface $S1^{x+}$ in Eq. S4 with the (known) activity of the bulk solution $S1_b^{x+}$:

$$a_{S1^{x+}} = a_{S1_b^{x+}} \cdot exp(-\frac{e\Psi_0}{kT}) \quad \text{(S6)}$$

Therefore, we may derive an explicit relationship between the bulk $S1^{x+}$ activity $a_{S1_b^{x+}}$ and the surface potential $\Psi_0$:

$$a_{S1_b^{x+}} = K_d \cdot exp(\frac{e\Psi_0}{kT}) \cdot (\frac{C_{DL}\Psi_0}{xeN_s - C_{DL}\Psi_0}) \quad \text{(S7)}$$

The mechanism can be modeled similarly in case of $ACE2 - S1$ complex system.